\documentclass[prl,twocolumn,preprintnumbers,amsmath,amssymb]{revtex4-1}
\usepackage{latexsym,graphicx,amssymb,amsmath,mathrsfs}
\usepackage{setspace,bm}
\usepackage[breaklinks, colorlinks=true, pdfstartview=FitV, linkcolor=red, citecolor=blue, urlcolor=blue]{hyperref}
\usepackage{amsmath}
\usepackage[usenames]{color}
\usepackage{latexsym}
\usepackage{epstopdf}
\usepackage{mathtools}

\newcommand\beq{\begin{equation}}
\newcommand\eeq{\end{equation}}
\newcommand\beqa{\begin{eqnarray}}
\newcommand\eeqa{\end{eqnarray}}

\newcommand\bq{{\bf q}}
\newcommand\br{{\bf r}}
\newcommand\bp{{\bf p}}

\usepackage[normalem]{ulem}

\begin{document}

\preprint{KUNS-2719, YITP-18-20}

\title{Anomalous Hall effect in dense QCD matter}

\author{Toshitaka Tatsumi}
\email[]{tatsumi@ruby.scphys.kyoto-u.ac.jp}
\affiliation{Department of Physics, Kyoto
University, Kyoto 606-8502, Japan}

\author{ Ryo Yoshiike}
\email[]{yoshiike@ruby.scphys.kyoto-u.ac.jp}
\affiliation{Department of Physics, Kyoto
University, Kyoto 606-8502, Japan}

\author{Kouji Kashiwa}
\email[]{kouji.kashiwa@yukawa.kyoto-u.ac.jp}
\affiliation{Yukawa Institute for Theoretical Physics,
Kyoto University, Kyoto 606-8502, Japan}

\begin{abstract}
In this letter, we investigate the anomalous Hall effect in dense
 QCD matter.
When the dual chiral density wave which is the spatially modulated
 chiral condensate appears in the medium,
 it gives rise to two Weyl points to the single-particle energy-spectrum and then
 the anomalous Hall conductivity becomes nonzero.
Then, dense QCD matter is analogous to the Weyl semimetal.
 The direct calculation of the Hall conductivity by way of Kubo's linear
 response theory gives the term
 proportional to the distance between the Weyl points.
Unlike the Weyl semimetal, there appears the additional contribution induced
 by axial anomaly.
\end{abstract}

\maketitle

\paragraph*{Introduction:}
It has been recently conceived that quantum chromodynamics (QCD) and
condensed matter physics share several important properties in phase
transition phenomena such as the
spontaneous symmetry breaking, Higgs mechanism and topological
structure of matters.
In this letter, we clarify deep relations between QCD and Weyl semimetal
from the topological viewpoint.

Understanding of the QCD phase structure 
at finite temperature ($T$) and baryon chemical potential ($\mu$)
is one of the important subjects in nuclear, hadron,
and elementary particle physics. 
It should be also interesting in the light of condensed matter physics.
One promising approach to access the QCD phase diagram
is the lattice QCD simulation
which is a powerful gauge-invariant approach to
investigate non-perturbative properties.
The lattice QCD simulation, however,
has the well-known sign problem at finite $\mu$.
Thus, one can only reach the region, $\mu/T < 1$,
even if we utilize several methods to circumvent the sign problem;
see Ref.~\cite{deForcrand:2010ys} as an example.
Therefore, several effective models haven been used to investigate the
QCD phase structure, qualitatively.

Recently, the chiral condensate with the spatial modulation
attracts much attention in QCD.
The inhomogeneous chiral phase (iCP) is characterized by the generalized order parameter,
\begin{align}
M \equiv   \langle {\bar \psi}\psi\rangle
       + i \langle{\bar \psi}i\gamma_5\tau_3 \psi\rangle
  = \Delta(\br) e^{i\theta(\br)}
\end{align}
for $SU(2)_L\times SU(2)_R$, where $\langle {\bar \psi} \psi \rangle$ and
$\langle {\bar \psi} i\gamma_5 \tau_3 \psi \rangle$
represent the scalar and pseudo-scalar quark condensates with the quark
field $\psi$, respectively.
The amplitude or phase of the quark condensates can be spatially
modulating.
There have been studied various iCP structures by solving the
self-consistent equations within the effective model of QCD such as the
Nambu-Jona-Lasinio (NJL) model~\cite{Buballa:2014tba}.
The typical forms of inhomogeneous chiral condensates
with one-dimensional modulations
are the dual chiral density wave (DCDW)~\cite{Nakano:2004cd}
and the real kink crystal (RKC)~\cite{Basar:2009fg,*Nickel:2009wj};
$M(z)= \Delta e^{iqz}$ for DCDW, and $M(z)
 = \frac{2 \Delta \sqrt{\nu}}{1+\sqrt{\nu}} \mathrm{sn}
   \Bigl( \frac{2 \Delta z}{1+\sqrt{\nu}},\nu \Bigr)$ for RKC,
where $\Delta \in \mathbb{R}$ is a constant amplitude,
$q \in \mathbb{R}$ denotes a wave-number of the one-dimensional
modulation in the $z$ direction
and $\mathrm{sn}(x,\nu)$ denotes the Jacobi elliptic function
with modulus $\nu \in [0,1]$.

In this letter, we focus on the DCDW phase to see an interesting
topological aspect. In the recent paper Ferrer and de la Incera have
pointed out that anomalous transport should be present in the DCDW phase
under the external magnetic field,
by modifying the Maxwell equations, which they called axion
electrodynamics \cite{Ferrer:2015iop,*Ferrer:2016toh}:
They found that the system induces the anomalous dissipationless Hall
current and the anomalous electric charge density. Since its various
consequences should be important phenomenologically,
it is indispensable to carry out further investigations.

The quark eigenenergy can be then given by diagonalizing the Hamiltonian
within the mean-field approximation;
the spectrum is symmetric with respect to zero and the positive-energy
solutions render
\begin{equation}
 E_s({\bf p})
 =\sqrt{ E_0^2 + \frac{|\bq|^2}{4} +s \sqrt{(\bq\cdot\bp)^2 + m^2|\bq|^2} },
\label{sp}
\end{equation}
using the NJL model in the chiral limit, where $s=\pm$, $m=-2G\Delta$
and $E_0=\sqrt{p^2+m^2}$ here $G$ is the coupling constant of the
four-quark interaction; for example, see~Ref.\cite{Nakano:2004cd}.
In following, we set $\mathbf{q}=(0,0,q)$.

\begin{figure}[b]
 \centering
 \includegraphics[width=0.33\textwidth]{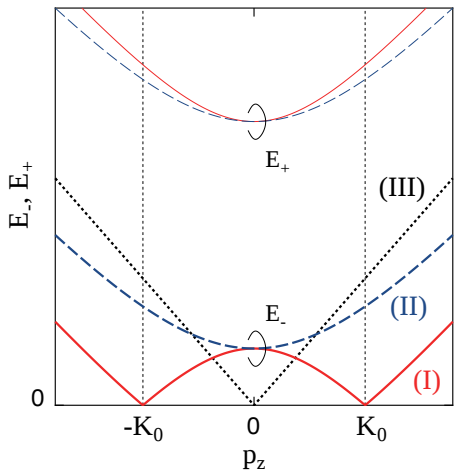}
 \caption{
 The single-particle spectrum as a function of $p_z$ with $p_\perp=0$.
 The cases (I), (II) and (III) represent the situation with
 $m<q/2$, $m>q/2$ and $m=q=0$, respectively.
 The thick and thin lines denote $E_-$ and $E_+$, respectively.}
 \label{Fig:S}
\end{figure}

The phase transition triggers off the instability of the Fermi surface
similar to nesting, so that the value of $q$ is much bigger than $m$,
$q/2\gg m$~\cite{Tatsumi:2004dx,Nakano:2004cd}.
Figure~\ref{Fig:S} shows a schematic view of the quark energy spectrum
for the cases with $m<q/2$ ((I)), $m>q/2$ ((II)) and $m=q=0$ ((III)), respectively.
The cases (I) and (II) are analogous to the Weyl semimetal and the spin-splitting 
insulator, respectively.
There appear two nodes called the Weyl points, $K_0=\sqrt{(q/2)^2-m^2}$ and
$-K_0$, in the case (I); for example see Ref.~\cite{murakami2007phase}.
The Dirac Hamiltonian can be well approximated by the $2\times
2$ Weyl Hamiltonian in the vicinity of each Weyl point. The
monopole-anti-monopole pair appears at the Weyl points and it gives rise
to topological effects.
The valence band is the Dirac sea and the conduction band
corresponds to the Fermi sea in the DCDW state.
Since the chemical potential is nonzero, we can regard the DCDW state as a
Weyl metal.

The Weyl semimetal is a topological material in three dimensions and
recently attracts much attention experimentally and theoretically
in condensed matter physics; for example, see
Ref.~\cite{yan2017topological,Armitage:2017cjs,sekine2016chiral}.
It is known that we can expect anomalous Hall effect (AHE) in the Weyl
semimetal. Applying the electric field $E$ along the $y$ axis, the
electric current density along the $x$ axis  $\langle j_x\rangle_E$ can
be measured: the anomalous Hall conductivity is then calculated as
\begin{align}
 \sigma_{xy} \equiv \frac{\langle j_x\rangle_E}{E}&= \frac{e^2}{(2\pi)^2} d_{\rm WP},
 ~~~\sigma_{ii}=\sigma_{xz}=\sigma_{zx}=0,
\end{align}
where $d_{\rm WP}=2K_0$ is the distance between the Weyl points, $e$
denotes the elementary charge and
$i=x,y,z$~\cite{wan2011topological,*yang2011quantum,liu2017giant}.
AHE can be expected, e.g., by doping  the
magnetic impurities in the Weyl semimetal. Correspondingly, DCDW bears a
ferromagnetic property \cite{Yoshiike:2015tha};
when we evaluate magnetization in response to the external magnetic
field $B$, we can see it survives nonzero in the limit $B\rightarrow 0$,
and the finite residual is given by the odd function of the wave number
$q$. Thus we may expect a similar phenomenon in the DCDW phase.

In this letter we evaluate the anomalous Hall conductivity by using the
Thouless-Kohmoto-Nightingale-den Nijs (TKNN) formula \cite{Thouless:1982zz}.
We start from the Dirac Hamiltonian with DCDW and calculate the
anomalous Hall conductivity by way of Kubo's linear-response theory.
We shall see that the anomalous Hall current is actually induced in the
DCDW phase because of the existence of Weyl points.
We also discuss its relation to axial anomaly in medium
and its implications of the anomalous transport properties of  dense
QCD matter.

\paragraph*{Anomalous Hall conductivity:}

We start from the two-flavor Dirac Hamiltonian with the DCDW;
\begin{align}
{\cal H} &= \left(
    \renewcommand{\arraystretch}{1.6}
    \begin{array}{ccc}
      \sigma \cdot (\mathbf{p}-\mathbf{q}\tau_3/2) & m \\
      m & -\sigma \cdot (\mathbf{p}+\mathbf{q}\tau_3/2)
    \end{array}
    \right),
\label{Eq:DM}
\end{align}
where $\sigma$ and $\tau$ represent Pauli matrices for the
spin and flavor spaces, respectively.
The positive single-particle spectrum is then given by
\begin{align}
 E_s ({\bf p})
 &= \sqrt{ p_\perp^2 
         + \Bigl( \sqrt{ m^2 + p_z^2} + s\frac{q}{2} \Bigr)^2 },
\end{align}
and the negative spectrum does $-E_s(\mathbf{p})$.
When $m$ and $q$ are non-zero, the spectrum split into two
branches and then we have two Weyl points.
The corresponding eigenspinors are
\begin{align}
 u_s ({\bf p}) &=
 N_s \left(
    \renewcommand{\arraystretch}{1.6}
    \begin{array}{ccc}
      \psi^{(s)}_L ({\bf p}) \\
      \psi^{(s)}_R ({\bf p})
    \end{array}
 \right),
 \nonumber\\
 v_s ({\bf p}) &= u_s ({\bf p}) \Bigl|_{E_s({\bf p}) \to - E_s({\bf p})}
\end{align}
where each component is expressed as
\begin{align}
 \psi_R^{(s)} ({\bf p}) &=
 \left(
    \renewcommand{\arraystretch}{1.6}
    \begin{array}{ccc}
      \cfrac{{\tilde E_s}({\bf p})}{p_\perp^2} p_-\\
      -1
    \end{array}
 \right),
 \nonumber\\
 \psi_L^{(s)} ({\bf p})
  &= - \frac{m}{{\tilde \epsilon_0}(p_z)} \sigma_3 \psi_R^{(s)} ({\bf p}),
\end{align}
with $p_\pm = p_x \pm i p_y$ and $\mathbf{p}_\perp = p_x^2 + p_y^2$.
The normalization factor then becomes
\begin{align}
 N_s^2 &= \frac{{\bf p}_\perp^2 {\tilde \epsilon_0}(p_z)}{4 s
 \epsilon_0 (p_z) E_s ({\bf p}) {\tilde E_s} ({\bf p})} ,
\end{align}
where
\begin{align}
{\tilde E_s}({\bf p}) &= E_s ({\bf p}) - s \epsilon_0 (p_z) - \frac{q}{2},
 \nonumber \\
\epsilon_0 (p_z) &= \sqrt{p_z^2 + m^2},~~~~
{\tilde \epsilon_0}(p_z) = p_z + s \epsilon_0(p_z).
\end{align}
Then, the Berry connection for the positive energy states is defined via the momentum derivative as
\begin{align}
 a^{(s)}_i({\bf p})
 &= -i \Bigl\langle u_s ({\bf p}) \hspace{0.5mm}
       \Bigl| \frac{\partial}{\partial p_i}
       \Bigl| \hspace{0.5mm} u_s({\bf p}) \Bigl\rangle,
\end{align}
and its curvature becomes
\begin{align}
 b_{s,xy}({\bf p})
 &= \frac{\partial a_y^{(s)}({\bf p})}{\partial p_x}
  - \frac{\partial a_x^{(s)}({\bf p})}{\partial p_y}
 \nonumber\\
 &=  -i \Bigl[
        \frac{\partial u_s^\dag ({\bf p})}{\partial p_x}
        \frac{\partial u_s ({\bf p})}{\partial p_y}
        -
        \frac{\partial u_s^\dag ({\bf p})}{\partial p_y}
        \frac{\partial u_s ({\bf p})}{\partial p_x}
        \Bigr].
\end{align}
The anomalous Hall conductivity in the $3+1$ dimensional system is then expressed as
\begin{align}
\sigma_{xy} &= e^2 \sum_{s} \int \frac{d p_z}{2\pi} \int \frac{d^2{\bf p}_\perp}{(2 \pi)^2}
  \Bigl[ b_{s, xy}({\bf p}) f(E_s({\bf p})) \Bigr],
\label{ahc}
\end{align}
where $f(E)$ is the Fermi-Dirac distribution function and
$e^2$ comes from the cancellation  due to different directions of the
wave vector for $u$ and $d$ quarks,
$N_\mathrm{c}(e_\mathrm{u}^2-e_\mathrm{d}^2)$
with $(e_\mathrm{u},e_\mathrm{d})=(2e/3,-e/3)$ and $N_\mathrm{c}=3$.
After few straightforward calculations,
the Berry curvature is finally expressed as
\begin{align}
 b_{s,xy} ({\bf p})
 &= -\frac{1}{2 E_s^3 }
    \Bigl(s\epsilon_0 + \frac{q}{2} \Bigr).
\label{ber}
\end{align}
When we evaluate contributions from the negative energy-spectrum,
$E_s$ should be replaced with $-E_s$ in Eqs.~(\ref{ahc}) and (\ref{ber}).
It should be noted that
$b_{s,xz}$ and $b_{s,yz}$ become the odd
function for $p_x$, $p_y$ or $p_z$, respectively.
Therefore, we have $\sigma_{xz}=\sigma_{yz}=0$.

Since $\mu$ is finite in the DCDW phase, the Hall conductivity $\sigma_{xy}$ consists of  two parts, $\sigma_{xy}=\sigma_{xy}^{\rm Dirac}+\sigma_{xy}^{\rm Fermi}$, where $\sigma_{xy}^{\rm Dirac}$ and $\sigma_{xy}^{\rm Fermi}$ are the contributions from the Dirac sea and the Fermi sea, respectively.

First, we consider the contribution of the Dirac sea at
$T=0$ because inclusion of the Fermi sea is straightforward.
The anomalous Hall conductivity can be expressed  as
\begin{widetext}
\begin{align}
 \sigma_{xy}^{\rm Dirac}
 &= e^2 \sum_{s} \int \frac{d p_z}{2\pi} \int \frac{d^2{\bf p}_\perp}{(2 \pi)^2}
    \Bigl[ \frac{1}{2 E_s^3 } \Bigl(s\epsilon_0 + \frac{q}{2} \Bigr)
 \Bigr] \nonumber\\
 &= \frac{e^2}{(2\pi)^2} \lim_{\Lambda \to \infty} \sum_{s}
    \int_0^{\Lambda_z} d p_z
    \left\{ \mathrm{sgn} \Bigl( s\sqrt{m^2 + p_z^2} + q/2 \Bigr)
          - \frac{s\sqrt{m^2 + p_z^2} + q/2}
                 { \sqrt{ \Lambda_\perp^2 + ( \sqrt{m^2 + p_z^2} + sq/2 )^2 } } \right\},
\label{Eq:AHC}
\end{align}
\end{widetext}
where $\mathrm{sgn}(\cdot)$ means the sign function and $\Lambda_\perp$ and
$\Lambda_z$ are related with $\Lambda$;
$\Lambda_\perp = \Lambda_z = \Lambda$ for the three-dimensional momentum
cutoff scheme
and $\Lambda_\perp = \sqrt{\Lambda^2-(\epsilon_0+sq/2)^2}$ and
$\Lambda_z = \sqrt{(\Lambda-sq/2)^2-m^2}$
for the energy cutoff scheme.
Unfortunately, the anomalous Hall conductivity depends on the cutoff
scheme;
\begin{align}
 \sigma_{xy}^\mathrm{Dirac} &= e^2 \times
 \begin{dcases}
  C, & m > \frac{|q|}{2} \\
  C + \mathrm{sgn}(q)\frac{d_{\rm WP}}{(2\pi)^2}, &m < \frac{|q|}{2} \\
 \end{dcases}
\end{align}
where $C$ denotes the cutoff dependent term.
In the three-dimensional momentum cutoff scheme, $C=-q/(6\pi^2)$.
In contrast, $C$ becomes $ -q/(2\pi)^2$ in the
energy cutoff scheme.
Such cutoff dependence is well known in the study of the Weyl
semimetal~\cite{Grushin:2012mt,Goswami:2012db}; we must impose a physical condition
to fix the cutoff dependence.
In the physical situation $C$ should be vanished because the
anomalous Hall conductivity should be zero for the insulator ($m>|q|/2$)
and gives Eq.~(3) for the Weyl semimetal ($m<|q|/2$) .
On the other hand, another criterion is needed to remove ambiguity for $C$,
when DCDW appears in QCD; the condition $m<|q|/2$ is always satisfied in the DCDW phase and
the massless limit ($m\rightarrow 0$) implies the disappearance of DCDW.
If $C$ is vanished, AHE remains in the normal quark phase and then the
nonzero anomalous Hall conductivity is unphysical.
Thus $\sigma_{xy}^{\rm Dirac}\rightarrow 0$ as $m\rightarrow 0$. This is
achieved in the gauge invariant regularizations such as the proper-time
method or the heat-kernel method \cite{Tatsumi:2014wka}.
The energy cutoff scheme can give the same answer in the present case,
\begin{align}
 \sigma^\mathrm{Dirac}_{xy} &= e^2 \times
 \begin{dcases}
  -\frac{q}{(2\pi)^2}, & m > \frac{|q|}{2} \\
  -\frac{q}{(2\pi)^2}
  +\mathrm{sgn}(q) \frac{d_{\rm WP}}{(2\pi)^2}, &m < \frac{|q|}{2}. \\
 \end{dcases}
\label{Eq:AHE-2}
\end{align}
This is our main result in this letter.
Note that only the lower case is realized in the DCDW phase to give the anomalous Hall current,
\begin{align}
 {\bf j}_{\rm AHE}
 &=\frac{e^2}{(2\pi)^2}\left(1-\frac{d_{\rm WP}}{|q|}\right){\bf q}\times {\bf E}.
\end{align}
We shall see that the first term stems from the effect of axial anomaly.
In the $m\to 0$ limit, AHE disappears which
means that axial anomaly exactly eliminates the anomalous Hall current
in QCD.

\begin{figure}[b]
 \centering
 \includegraphics[width=0.33\textwidth]{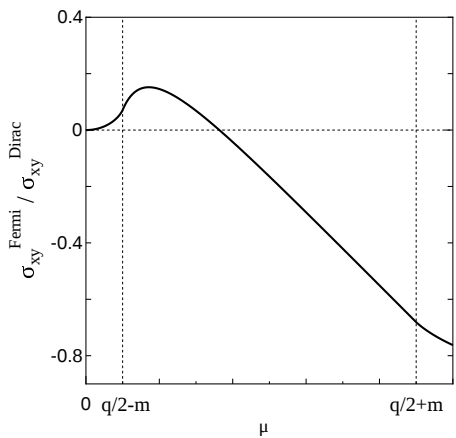}
 \caption{
 The $\mu$-dependence of $\sigma_{xy}^\mathrm{Fermi}$ with fixed
 $m$ and $q$.
 }
 \label{Fig:AHC}
\end{figure}

Next, we take into account the Fermi-sea contributions.
The Fermi-sea contributions for the cases
(a) $\mu<q/2-m$, (b) $q/2-m < \mu < q/2+m$ and (c) $q/2+m<\mu$ become
\begin{widetext}
\begin{align}
 \sigma_{xy}^{\rm Fermi}
    &= \frac{e^2}{(2\pi)^2} \Bigl[
      \frac{1}{2\mu} \Bigl(\mu + \frac{q}{2} \Bigr)^2 \sin \theta_+
    - \frac{1}{2\mu}\Bigl(\mu - \frac{q}{2} \Bigr)^2 \sin \theta_-
    - \frac{m^2}{4\mu} \ln
      \frac{(1+\sin\theta_+)(1-\sin\theta_-)}{(1-\sin\theta_+)(1+\sin\theta_-)}
    - d_\mathrm{WP}\Bigr] , 
    %
 \label{Eq:AHC_F}
\end{align}
\end{widetext}
where
$\sin \theta_+ = \sqrt{1-m^2/(\mu+ q/2)^2}$, and $\sin \theta_ -= \sqrt{1-m^2/(\mu- q/2)^2}$ for (a) or (c)  and $\theta_-=0$ for (b).
For the Fermi-sea contribution, we do not need to fix the cutoff scheme
because the step function acts as the natural cutoff.
Figure~\ref{Fig:AHC} shows the $\mu$-dependence of
$\sigma_{xy}^\mathrm{Fermi}$ with fixed $m$ and $q$.
It correctly approaches to zero with $\mu \to 0$.
In the realistic situation, $m$ and $q$ changes as functions of $\mu$ , 
and thus we must carefully extract the effect of  $\sigma_{xy}^\mathrm{Fermi}$.

By taking $m \to 0$ limit, we have $\sigma^\mathrm{Fermi} \to 0$ and
thus this expression is suitable from the viewpoint of the
normal quark matter behavior.

\paragraph*{Discussion:}
One may wonder whether our result can be derived from considerations of
axial anomaly \cite{Zyuzin:2012tv}.
Actually, using Fujikawa's method, Ferrer and de la Incera have
discussed AHE and axion electrodynamics in the DCDW phase in the
presence of the background electromagnetic field  \cite{Ferrer:2015iop,*Ferrer:2016toh}.
Since the Dirac Hamiltonian (4) can be obtained by the local chiral
transformation on the quark field,
$\psi\rightarrow {\rm exp}(-i\tau_3\gamma_5\theta/2)\psi$, with
$\theta={\bf q}\cdot{\bf  r}$ from the original one, the anomaly term
should appear in the action in the presence of the electromagnetic
field,
$S_\mathrm{anom}=-(e^2/16\pi^2)\int d^4 x \theta ~F_{\mu\nu}{\tilde F}^{\mu\nu}$.
The variation of $S_\mathrm{anom}$ with respect to $A_i$ gives the anomalous Hall current,
$j_\mathrm{anom}^i=-\delta S_{\mathrm anom}/\delta A_i$;
\begin{align}
 j^i_\mathrm{anom} = \frac{e^2}{4\pi^2}\nabla\theta\times {\mathbf E}
             = \frac{e^2}{4\pi^2} {\mathbf q}\times {\mathbf E},
\end{align}
from which the anomalous Hall conductivity  reads as
$\sigma_{xy}=-e^2/(4\pi^2)q$. Note that this expression is incorrect in
comparison with Eq.~(\ref{Eq:AHE-2}). Instead, we can see that the first term in Eq.~(\ref{Eq:AHE-2}) comes from axial anomaly.

We can see the similar situation in the evaluation of anomalous quark
number, which is brought about by spectral asymmetry of the quark fields.
It has been shown that spectral asymmetry does not necessarily gives the same
result as the one given by axial anomaly
\cite{Tatsumi:2014wka}: it coincides with each other only in the case,
$q/2\ll m$ which is corresponding to the situation that the Weyl points
disappear.
This means that we can not reach the total amount of the anomalous quark number
by only using Fujikawa's method.
To evaluate the anomalous transport coefficients correctly, we must use
Kubo's linear response theory.

\paragraph*{Summary:}
In this letter, we have considered the anomalous Hall effect (AHE) in
 dense QCD matter.
At finite density, we can expect that the spatial inhomogeneity appears
in the chiral condensate as the dual chiral density wave (DCDW) and the
DCDW can lead topologically nontrivial properties to QCD such as AHE.

Starting from the Dirac Hamiltonian with  DCDW, we can evaluate the
Berry connection and its curvature from the one-particle spectrum and
corresponding eigen-functions.
Then, the anomalous Hall conductivity can be obtained by way of the TKNN
formula based on Kubo's linear response theory.
It is found that  dense QCD matter can exhibit AHE when DCDW appears.
Depending on the strength of the amplitude ($m$) and the phase
($q$) of DCDW,
the anomalous Hall conductivity shows the different functional form but
it is always nonzero with $q \neq 0$ except for the $m \to 0$ limit.

Interestingly, we cannot obtain whole amount of the anomalous Hall
conductivity by using Fujikawa's method,
which has been used to estimate the anomalous transport properties in
QCD under external
electromagnetic fields.
To correctly estimate the anomalous transport in the system, we should
calculate the anomalous Hall conductivity by way of Kubo's linear response
theory.

One of the important consequences of the anomalous Hall current is 
the modification of the Maxwell equations. It should affect the electromagnetic 
transport properties of dense matter inside neutron stars by way of the 
magnetohydrodynamics (MHD) \cite{Tajima2002Plasma}. 

Our calculation can be easily extended to the magnetic DCDW phase in the
presence of the magnetic
field by using the Streda formula \cite{streda1982theory}. The
expression of the anomalous Hall conductivity coincides with
Eq.~(\ref{Eq:AHE-2}). Details will be presented in our forthcoming
paper~\cite{tat2}.

Finally it would be worth mentioning that Weyl semimetals provide us
with a laboratory to study dense QCD.
We hope future observations about Weyl semimetals such as AHE or induced
charge can check or verify the ideas obtained in the study of dense QCD
matter.

\begin{acknowledgments}
We thank K. Nomura and Y. Kikuchi for useful discussions and information about the Weyl semimetal.
This work is partially supported by Grants-in-Aid for Japan Society
 for the Promotion of Science (JSPS) fellows ~No.27-1814.
\end{acknowledgments}

\bibliography{ref.bib}

\end{document}